\documentclass{article}

\usepackage{spconf,amsmath,graphicx,epsfig}
\usepackage{amssymb,amsfonts, amstext}
\usepackage{epsfig}
\usepackage{epstopdf}
\usepackage[font=footnotesize]{caption}
\usepackage{array}
\usepackage{multicol}
\usepackage{multirow}
\usepackage{makecell}
\usepackage{verbatim} 

\title{Multimodal Signal Processing and Learning Aspects of Human-Robot Interaction for an Assistive Bathing Robot}
\name{A. Zlatintsi, I. Rodomagoulakis, P. Koutras, A. C. Dometios, V. Pitsikalis, C. S. Tzafestas, and P. Maragos
\thanks{This research work was supported by the EU under the project I-SUPPORT with grant H2020-643666.}}
\address{School of E.C.E., National Technical University of Athens, 15773 Athens, Greece \\ \small{Email: \{nzlat, irodoma, pkoutras, vpitsik, ktzaf, maragos\}@cs.ntua.gr, athdom@mail.ntua.gr}}
\begin{document}
\ninept
\maketitle
\begin{abstract}
We explore new aspects of assistive living on smart human-robot interaction
(HRI) that involve automatic recognition and online validation of speech and gestures in a natural interface, providing social features for HRI. We introduce a whole framework and resources of a real-life scenario for elderly subjects supported by an assistive bathing robot, addressing health and hygiene care issues. We contribute a new dataset 
and a suite of tools used for data acquisition and a state-of-the-art pipeline for multimodal learning 
within the framework of the I-Support bathing robot, 
with emphasis on audio 
and RGB-D visual streams. We consider privacy issues by evaluating the depth visual stream along with the RGB, using Kinect sensors. 
The audio-gestural recognition task on this new dataset yields up to 84.5\%, while the online validation of the I-Support system on elderly users accomplishes up to 84\% when the two modalities are fused together. The results are promising enough to support further research in the area of multimodal recognition for assistive social HRI, considering the difficulties of the specific task. Upon acceptance of the paper part of the data will be publicly available. 
\end{abstract}
\begin{keywords}
human-robot communication, assistive HRI, multimodal dataset, audio-gestural command recognition, online validation with elderly users
\end{keywords}
%
\section{Introduction}
\label{sec:intro}

During the last decades, an enormous number of socially interactive robots have been developed constituting the field of Human-Robot Interaction (HRI) an actual motivating challenge. This challenge has become even greater, due to their relocation outside the lab environment and into real use cases. Robotic agents are nowadays able to serve in various new roles, assisting with every day tasks involved in the caring of elderly or children. The aim of such systems is to shift part of the care burden from healthcare professionals and to support individuals to manage and to take better care of their own health, to enable independent living and to improve their life quality. 

The evolution of the robotic industry towards this end have given rise to new social, ethical, privacy challenges etc., and have led HRI research to extend into many research areas \cite{GoSc07}. One such area concerns the development of multimodal interfaces, required to facilitate natural HRI, including visual RGB/Depth and audio input, for multimodal human-robot interactions. 
The concept behind these systems 
is to design robots and interaction techniques that will enhance the communication making it natural and intuitive, so as the robots are able to understand, interact and respond to human intentions intelligently; for a review, we refer the reader to \cite{KSKC14,JCJ+14,EFG+16,RWBM15,KoWe16,RKP+16,ZRP+17}.

Arguments about robotic systems in health and hygiene care \cite{KCJC10} include: a) the economic advantages in supporting the growing 
elderly population, b) the increased independence and life quality and c) the increased sense of comfort and privacy; features that are highly valued by the end-users \cite{openroboethincs}. 
However, domotic robots are equipped with cameras and microphones not only for communication but for monitoring and security, as well. So, an issue that has drawn much discussion is whether such sensor infrastructure 
could infridge on the right to privacy, especially when an elderly person's mental health deteriorates.

\begin{figure}[!tb!]
\centering 
\includegraphics[ trim=0 0cm 0 0cm, clip, width=\linewidth]{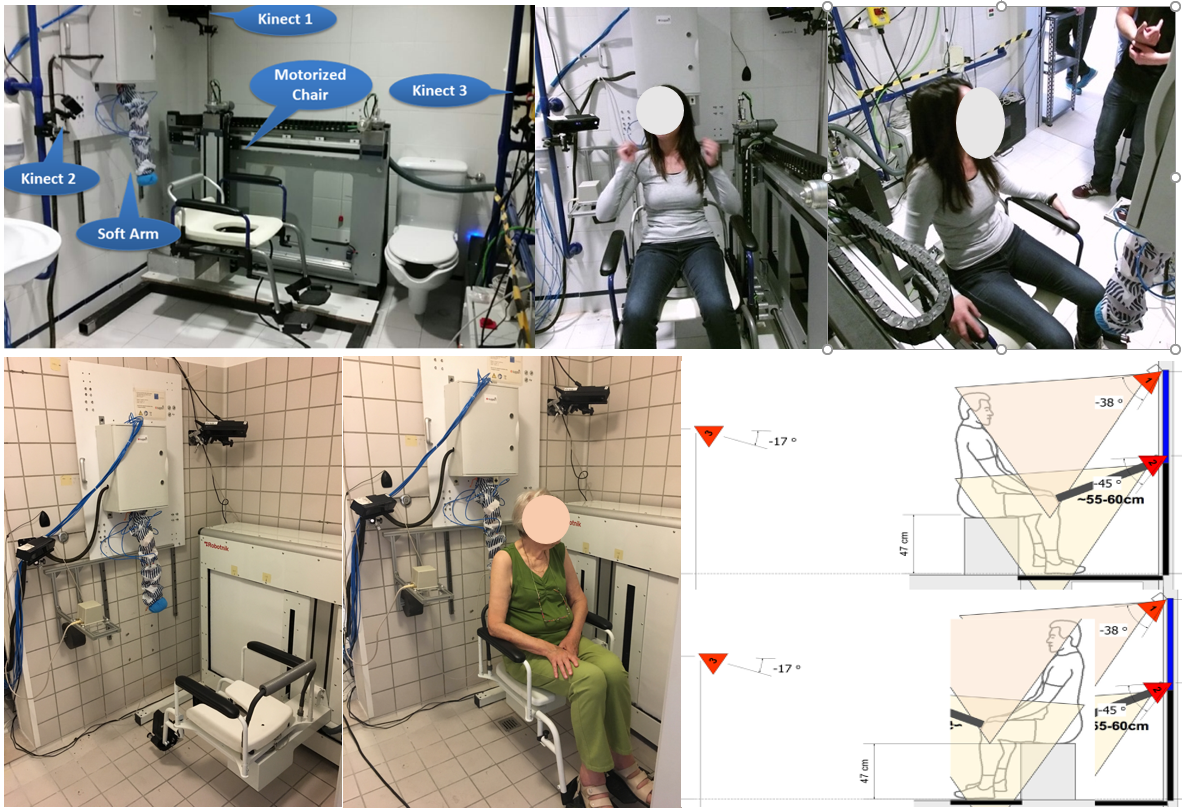}
\vspace{-0.6cm}
\caption{The I-Support automatic bathing environment. Setup of the Kinect sensors as installed in the two validation sites, FSL (top) and Bethanien (bottom) hospitals.}
\label{fig:onval}
\vspace{-0.75cm}
\end{figure}

In this paper, we contribute an experimental framework and system on social human-robot interaction via a rich HRI set of resources including a domain specific dataset and automatic machine learning tools. These are applied on a real-life use case of an assistive bathing robot, which helps elderly to perform and complete bathing tasks, towards independent living and improved life quality. The domain specific development dataset that is introduced includes audio commands, gestures -- an integral part of human communication \cite{McNe96} -- and co-speech gesturing data, which is still quite limited in HRI \cite{Cass98}. Our goal is to enhance the communication making it natural, intuitive and easy to use, thus, enhancing it wrt social aspects. Additionally, we present an automatic multimodal recognition system, based on state-of-the-art signal processing techniques and pattern recognition algorithms that are able to model audio-visual data, i.e., speech and RGB-D, so as to recognize, in an online manner, multimodal data addressing aspects in assistive smart HRI. We present promising results both when evaluating the two streams independently (offline evaluation) and when validating online the human-robot interaction between the I-Support system and the elderly end-users.  
%

\section{I-SUPPORT: a robotic shower system}
\label{case_study}
\vspace{-0.3cm}
\begin{figure}
\centering
\includegraphics[trim=0 0cm 0 0cm, clip, width=\linewidth]{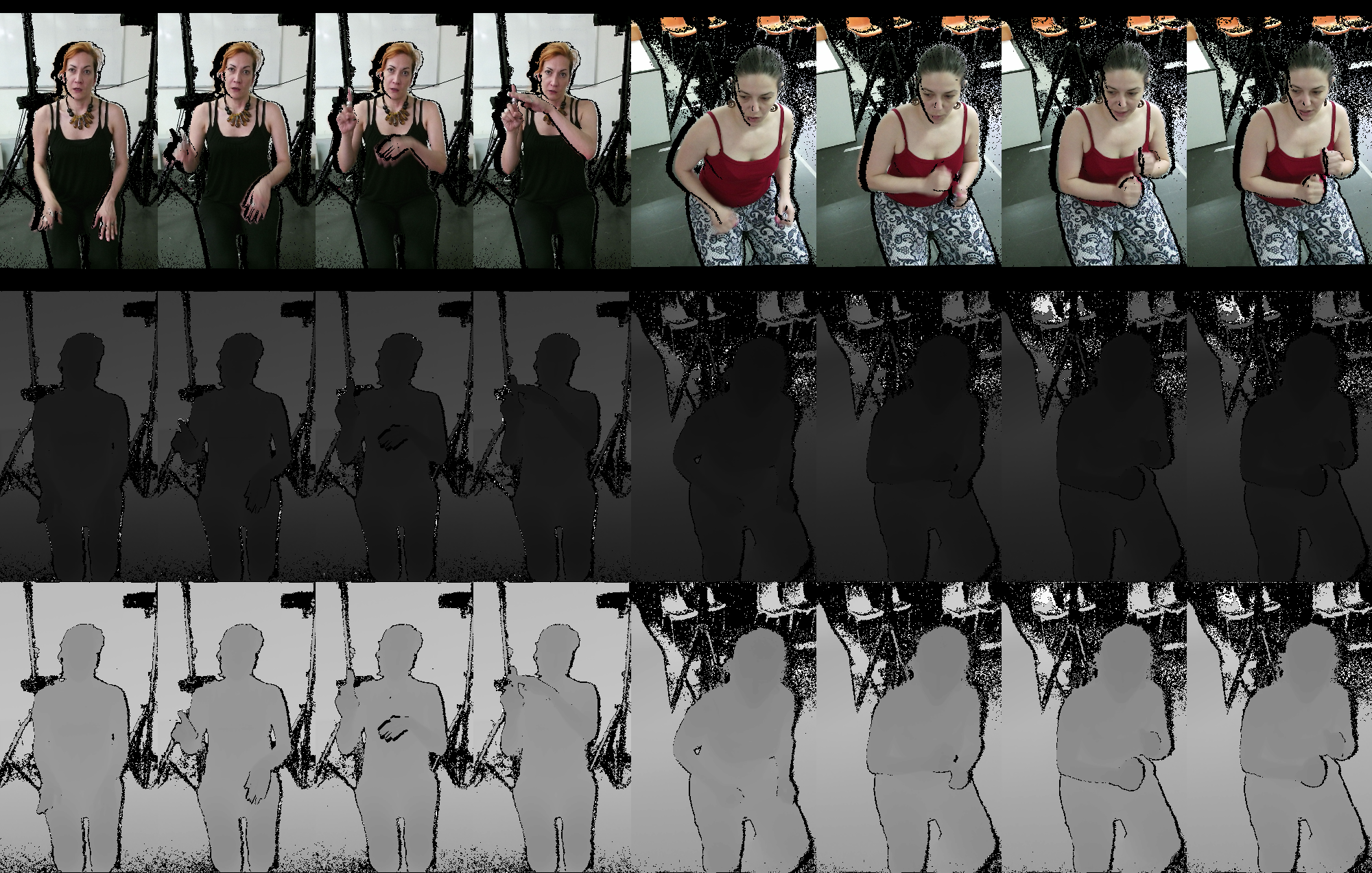}
\includegraphics[trim=0 0cm 0 0cm, clip, width=\linewidth]{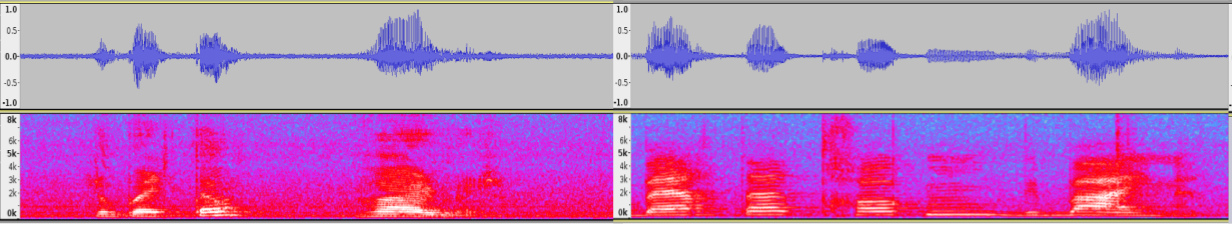}
\vspace{-0.6cm}
\caption{Data streams acquired by sensors \#1 and \#2: RGB (top), depth (2nd row) and log-depth (3rd row) frames from a selection of gestures (``Temperature Up", ``Scrub Legs"), 
accompanied by the corresponding German spoken commands -- waveforms (4th row) and spectrograms (bottom row) -- ``W\"armer", ``Wasch die Beine". 
}
\vspace{-0.6cm}
\label{fig:data}
\end{figure}

The goal of the I-Support\footnote{\scriptsize{http://www.i-support-project.eu/}} project is to develop a robotic bathing system that will enable independent living and improved life quality for the elderly. The core system functionalities identified as important from a clinical perspective (taking into account various impairments, limitations and user requirements) are the tasks for bathing the distal region (legs) 
and the back region 
\cite{WeHa16}. The experimental prototype, for natural human-robot communication, consists of 
three Kinect sensors, as shown in Fig.~\ref{fig:onval}, for 3D pose reconstruction of the scene (user and robot) and identification of user gestures, and an audio system including 8 distributed condenser microphones. The aforementioned sensors are also used to record, challenging multimodal data for modeling the user-robot interaction, including 
audio-gestural commands related to the specific tasks. 

Even though gestures, as means of communication, could be questioned, since elderly have to 
perform and remember them regardless possible impairments or their cultural background, the reason for using them is manifold. ``\textit{Gestures are an integral part of language as much as are words, phrases, and sentences –-
gesture and language are one system}'' \cite{McNe96}. It is also widely accepted that gestures aid human communication significantly by elaborating upon and enhancing the
information, which is orally co-expressed~\cite{Host11,McNe96}, while almost 90\%
of spoken utterances in descriptive conversations are accompanied by
speech-synchronized gestures \cite{Nobe00}, which occur similarly for
speakers of more than 20 cultures 
\cite{McNeil+10}.  

\textbf{Multi-view system architecture: }
Considering various constraints (i.e., the size of the bath cabin, the size and
the placement of the chair and the soft-arm robot base) we defined the setup for the three Kinect sensors, so as to be able to capture the necessary information for both washing tasks, i.e., information of the user’s back and legs as well as the hand gestures performed for the communication with the robot. Specifically, two of the Kinect sensors (Kinect sensors \#1 and \#2) were placed inside the bath cabin, in order to capture the legs and the back of the user during the
different tasks, while a third camera (Kinect sensor \#3) was placed outside the
cabin, in order to capture the gestures performed by the user during the task
washing the back, see Fig.~\ref{fig:onval}. 
Specifically, during the task washing the legs Kinect sensor \#2 records the legs of the user (including registered RGB and depth in SD resolution), for body pose estimation and visual tracking of the robot; while 
sensor \#1 is used by the gesture and action recognition module. Except from the streams in SD resolution sensor \#1 also records the color stream (RGB) in Full HD. 
During the task washing the back sensor \#1 records the back of the user (including RGB and depth in SD resolution), while sensor \#3 records the color stream Full HD, used for gesture recognition. 
\vspace{-0.3cm}
\section{System Description and offline evaluation }
\vspace{-0.3cm}
\textbf{Audio-gestural data acquisition: }
Using the proposed system architecture, we have collected an audio-gestural development dataset for the multimodal recognition task. Specifically, we have recorded visual data from 23 users while performing predefined gestures, and audio data from 8 users while uttering predefined spoken commands in German (the users were non-native German speakers, having only some beginner's course). The recorded gesture commands included certain variability, so as to be able to analyze them and henceforth to take care of factors such as intuitiveness and naturalness for the performance (how the users feel more comfortable to perform the various gestures), as well as to design a system that could recognize smaller or larger variations of the same command. 

The total number of commands for each task was: 25 and 27 gesture commands for washing the legs and the back, respectively, and 23 spoken commands --preceded by the keyword Roberta -- for the core bathing tasks, i.e., washing/scrubbing/wiping/rinsing the back or legs, for changing base settings of the system, i.e., temperature, water flow and spontaneous/emergency commands. A background model was also recorded, including generic motions or gestures that are actually performed by humans during bathing; so as to be able to reject out-of-vocabulary gestures (motions/actions) as background actions. For data collection and the simultaneous annotation of segments when the users were performing and/or uttering the spoken commands, we employed the graphical interface tool (GUI) described in \cite{kardaris2016platform}. Figure~\ref{fig:data} shows examples of the acquired data streams. Depth and log-depth are also explored and compared to RGB, targeting robust and privacy-aware visual processing.

\begin{figure}[!tb]
\centering
\includegraphics[scale=0.15]{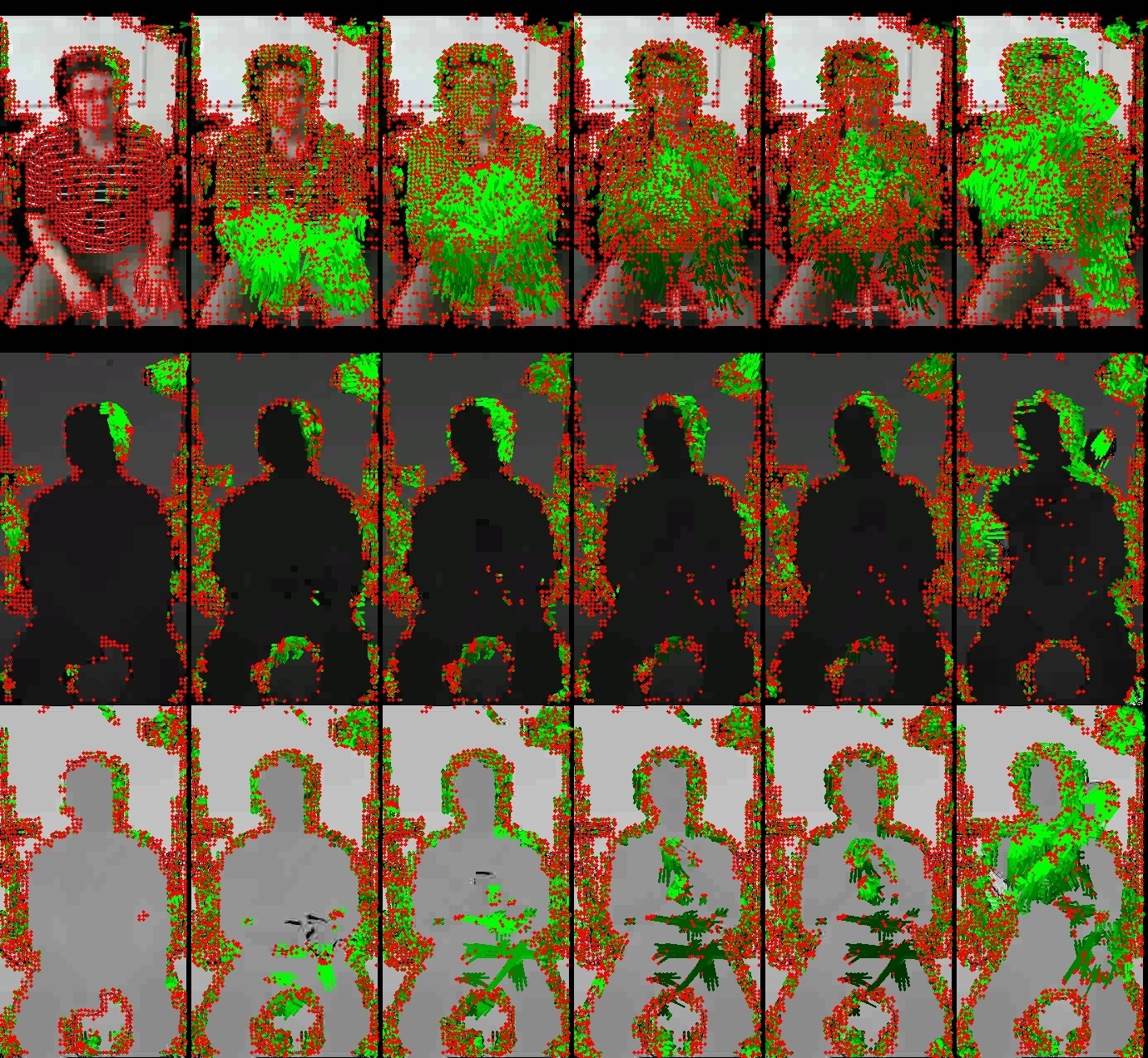}
\caption{Comparison of dense trajectories extraction over the RGB (top), depth (middle) and log-depth (bottom) clips of gesture ``Scrub Back".}
\vspace{-0.6cm}
\label{fig:dense}
\end{figure}

\textbf{Gesture classification experiments and results: }
Gesture recognition allows the interaction of the elderly subjects with the robotic platform through a predefined set of gestural commands. 
For this task, we have employed state-of-the-art computer vision approaches for feature extraction, encoding, and classification. Our gesture and action classification pipeline employs Dense Trajectories 
along with the popular Bag-of-Visual-Words (BoVW) framework. Dense Trajectories~\cite{wang2013action} has received attention, due to their performance on challenging datasets and the main concept consists of sampling feature points $n$ from each video frame on a regular grid and tracking them through time based on optical flow. Specifically the employed descriptors are: the Trajectory descriptor, HOG~\cite{wang_evaluation_2009}, HOF~\cite{laptev2008learning} and Motion Boundary Histograms (MBH)~\cite{wang_evaluation_2009}. As depicted in Fig.~\ref{fig:dense}, non-linear transformation of depth using logarithm (log-depth) enhances edges related to hand movements and leads to richer dense trajectories on the regions of interest, close to the result obtained using the RGB stream.

The features were encoded using BoVW and were assigned to $K=4000$ clusters forming the representation of each video. Afterwards, each trajectory was assigned to the closest visual word and a histogram of visual word occurrences was computed, yielding a sparse $K$-dimensional representation. Videos encoded with BoVWs were classified with non-linear SVMs using the $\chi^2$~\cite{wang_evaluation_2009} kernel, and different descriptors were combined in a multichannel approach, by computing distances between BoVW histograms as:
\vspace{-0.3cm}
\begin{equation}
K\left(\mathbf{h}_i,\mathbf{h}_j\right)=
\exp\left( -\sum_c \frac{1}{A_c} D\left(\mathbf{h}_i^c,\mathbf{h}_j^c\right)\right),
\label{eq:multichannel}
\vspace{-0.2cm}
\end{equation}
where $c$ is the $c$-th channel, i.e., $\mathbf{h}_i^c$ is the BoVW representation of the $i$-th video, computed for the $c$-th descriptor, and $A_c$  is the mean value of $\chi^2$ distances $D\left(\mathbf{h}_i^c,\mathbf{h}_j^c\right)$ between all pairs of training samples. 

The features were also encoded using VLAD (Vector of Locally Aggregated Descriptors) \cite{JDSP10}; first order statistics were encoded among features by computing differences between extracted features and the visual words. Videos encoded with VLAD were classified using linear SVMs and different descriptors were combined by concatenating their respective vectors. Since multiclass classification problems were considered, an one-against-all approach were followed and the class with the highest score was selected.

Table~\ref{tab:gres} presents average accuracy results (\%) for 25 and 27 gestures for the tasks washing the legs and the back, respectively, performed by 23 subjects. All features using the two encodings were evaluated and are shown for RGB data and log-depth (D) data. 
The combination of the proposed features (denoted with the label Comb.) performs consistently better for both evaluation setups and for both bathing tasks. Additionally, the motion related descriptors (HOF, MBH) yield better results than the static (HOG), since all employed gestures include motion of the whole arm rather than motion of the handshape. Regarding the two encodings, VLAD vector further improves the performance, since it encodes rich information about the visual words' distribution. Concerning the RGB vs. D evaluation, we note that RGB accomplishes better results than the log-depth (D). Concluding, we observe that the gesture classification system shows a rather good performance of about $83\%$ and $85\%$ (for VLAD and RGB) for the tasks washing the legs and the back, respectively. 

\begin{table}[!tb!]
\centering
\footnotesize{
\begin{tabular}{|l|c|c|c||c|c|} \cline{3-6}
           \multicolumn{1}{c}{}  &                                        & \multicolumn{2}{c||}{Task: Legs} & \multicolumn{2}{c|}{Task: Back} \\\hline
Feat. & Encoding                                  & RGB   & D   & RGB   & D  \\ \hline\hline
Traj.        & \multirow{5}{*}{BoVW}& 69.64& 60.52&  77.84 & 60.87         \\\cline{1-1} \cline{3-6} 
HOG          & &41.01 &53.34   &58.51 &57.14          \\\cline{1-1} \cline{3-6} 
HOF          & &74.15 &66.26   &82.92 &71.58          \\\cline{1-1} \cline{3-6} 
MBH          & &77.36 &65.31   &80.81 &65.73          \\\cline{1-1} \cline{3-6} 
Comb.        & &\textbf{80.88} &\textbf{74.41}  &\textbf{83.92} &\textbf{75.70}      \\\hline\hline
Traj.        & \multirow{5}{*}{VLAD}&69.22 &52.66   &  74.34&54.14  \\\cline{1-1} \cline{3-6}
HOG          & &49.86 &65.99   &      61.23&65.63          \\\cline{1-1} \cline{3-6} 
HOF          & &76.54 &72.88   &      83.17&78.07          \\\cline{1-1} \cline{3-6} 
MBH          & &78.35 &75.12   &      82.54&73.09          \\\cline{1-1} \cline{3-6} 
Comb.        & &\textbf{83.00} &\textbf{78.49} &       \textbf{ 84.54}&\textbf{81.18}          \\\hline
\end{tabular}}
\vspace{-0.2cm}
\caption{Average classification accuracy (\%) for the pre-defined gestures performed by 23 subjects. Results for the four features and their combination, using the two encodings are shown for RGB data and D (depth) data for the two bathing tasks. 
}
\vspace{-0.6cm}
\label{tab:gres}
\end{table}

\textbf{Spoken commands classification experiments and results: }
We have developed a spoken command recognition module~\cite{RKP+16} that detects and 
recognizes commands provided by the user freely, at any time, 
among other speech and non-speech events possibly infected by environmental 
noise and reverberation. We target robustness via a) denoising of the far-field signals, b) adaptation of the acoustic models, and c) combined command detection/recognition. Herein, the classification of pre-segmented commands is performed based on a task-dependent grammar of 23 German 
commands, 
using the signals captured by the central microphones (channel 2) of the  sensors \#1 and \#3 for the tasks washing the legs and the back. Leave-one-out experiments were conducted by testing classification on one subject after applying global MLLR adaptation of the acoustic models on the commands of the rest subjects, yielding average accuracies of $75.8\%$ and $67.6\%$ on 8 subjects for the two tasks.

\vspace{-0.3cm}
\section{Online validation with primary end-users}
\label{online}
\vspace{-0.3cm}

\textbf{Online A-G command recognition system: }The online A-G multimodal action recognition system that we have developed \cite{kardaris2016platform}, using the Robotic Operating System (ROS), enables the interaction between the user and the robotic arms so as to monitor, analyze and predict the user's actions, giving emphasis to the command-level speech and gesture recognition. Always-listening recognition is applied separately for spoken commands and gestures, combining at a second fusion level their results. The output is fed to the robot's controller and the predicted action or task is executed. The overall system comprises two separate sub-systems: (i)~the activity detector, which performs temporal localization of segments containing visual activity and (ii)~the gesture classifier, which assigns each detected activity segment into a class. The activity detector processes the RGB or the depth stream in a frame-by-frame basis and determines the existence of visual activity in the scene, using a thresholded activity ``score'' value. The gesture classification 
system is 
signaled at the beginning and at the end of the activity segments and herein, it processes 
and assigns them to one of the pre-defined categories.

\textbf{Validation studies with primary end-users: }We evaluated the online A-G recognition system, regarding its functionality and the human-robot interaction between the I-Support bathing robot and the primary end-users (elderly), using audio and audio-gestural commands. During the experiments, we simulated the two bathing scenarios (bathing the legs and back) at dry conditions, at two pilot sites: 1)~the Fondazione Santa Lucia (FSL) Hospital (Rome, Italy) and 2) the Bethanien Hospital (Heidelberg, Germany). Demanding acoustic and visual conditions were faced along with large variability in the performance of the audio-gestural commands by the participants, constituting the recognition task rather challenging. A statistically significant number of users, having various cognitive impairments, 
were selected by the clinical partners for the HRI experiments; thus, 25 (mean age$\pm$SD: 67.4$\pm$8.9 years) and 29 (mean age$\pm$SD: 81.4$\pm$7.7 years) patients were recruited on each site. 

The experimental protocol 
exhibited a variety of 7 audio or audio-gestural commands in Italian and German (see Table~\ref{tab:A-Gcommands}) in sequences that would simulate realistic interaction flows for both 
tasks. Table~\ref{tab:IDs} shows the sequence of the audio (A) and audio-gestural (A-G) commands as performed in the validation experiments. Prior to the actual testing phase, 
all commands were introduced to the participant by the clinical test administrator. During the 
experiment, the test administrator guided the participant on how to interact with the robot by showing the audio commands written on posters or the audio-gestural commands by performing them, and instructed him/her to simply read or mimic them. 
The administrator could also intervene whenever the flow was changed unexpectedly after a system failure. Additionally, a technical supervisor handled the PCs and annotated on-the-fly the recognition results of the system. 

\begin{table}[!tb]
\centering
\footnotesize{
\renewcommand{\arraystretch}{0.6}
\begin{tabular}{|c|c|c|}
\hline
\multicolumn{3}{|c|}{\textbf{Vocabulary}} \\ \hline
\textbf{English}  & \textbf{Italian} & \textbf{German}  \\ \hline
Wash legs          & Lava le gambe & Wasch meine Beine       \\ \hline
Wash back          & Lava la schiena & Wasch meinen R\"ucken      \\ \hline
Scrub back         & Strofina la schiena & Trockne meinen R\"ucken      \\ \hline
Stop (pause) & Basta  & Stop \\ \hline
Repeat (continue)& Ripeti & Noch einmal            \\ \hline
Halt  & Fermati subito & Wir sind fertig \\ \hline
\end{tabular}}
\vspace{-0.3cm}
\caption{The audio-gestural commands that were included in the two bathing scenarios. All commands were preceded by the keyword Roberta.}
\vspace{-0.4cm}
\label{tab:A-Gcommands}
\end{table}

\begin{table}[!tb]
\centering
\footnotesize{
\renewcommand{\arraystretch}{0.6}
\begin{tabular}{|c|c|c||c|c|}
\hline
& \multicolumn{2}{c||}{\textbf{Distal Region}} & \multicolumn{2}{c|}{\textbf{Back Region}}  \\ \hline
\textbf{ID}&\textbf{Command }&\textbf{Modality }&\textbf{Command} &\textbf{Modality }\\ \hline
1& Wash Legs &A & Wash Back &A-G \\ \hline
2& Stop &A &Halt&A-G \\ \hline
3& Repeat &A & Scrub Back&A-G \\ \hline
4& Halt &A & Stop &A-G \\ \hline
5&Wash Legs&A-G & Repeat &A-G \\ \hline
6& Halt &A-G & Halt &A-G  \\ \hline
7& Halt &A-G   & Halt &A-G \\ \hline
\end{tabular}}
\vspace{-0.3cm}
\caption{The sequence of Audio (A) and Audio-Gestural (A-G) commands performed by the participants in the validation experiments.}
\vspace{-0.6cm}
\label{tab:IDs}
\end{table}

The Kinect sensors and the microphones were installed in the bathrooms of the two hospitals 
according to the setup described in Sec.~\ref{case_study}; incorporating 
some adjustments regarding their 
positions and angles depending on the available space of each room. 
In addition, the A-G recognition system's grammars and functionalities were adapted to the specific bathing tasks delivering recognition results as ROS messages to the system’s finite state machine (FSM) that a) decided the action to be taken after each recognized command, b) controlled the various modules and c) managed the dialogue flow by producing the right audio feedback to the user. The individual speech and gesture recognition hypotheses were combined using a late fusion scheme encoding the inter-modality agreements and following the ranking rule: ``if the best speech recognition hypothesis is among the 2-best gesture hypotheses, then it is announced as the multi-modal result'' \cite{kardaris2016platform}. 

%

%
%

\textbf{Online Recognition Results: }
Multimodal recognition was evaluated in terms of 1) Multimodal Command Recognition Rate (MCRR): MCRR= \# of commands correctly recognized by the system $/$ \# of commands correctly performed by the user, 2) accuracy and 3) user performance/learning rate. 
Table~\ref{fig:mcrr} shows the obtained MCRR (\%) and accuracy results (\%), which are up to 84\% and 80\% for both washing tasks (averaged across 25 and 29 users). The result deviations for the two tasks are due to the different setups of the sensors, as emerged because of the space of the bathrooms. 
Gesture recognition was expected more challenging while bathing the legs, due to occlusions of the hands with the robot and/or the chair. 
The users experienced only a limited amount of false alarms (3 in total as measured at FSL), which were considered annoying, since the system triggered a response without an ``actual'' input. Regarding the user performance, the participants performed successfully the spoken commands (over 90\% accuracy), while the average performance of gestures was satisfactory (between 70\% to over 80\%) after the quick training provided by the 
administrator. As seen in Table~\ref{fig:mcrr}, the participants in Bethanien were less capable in performing the commands and this is also the reason of the lower results in MCRR and accuracy.  Finally, we have to mention that the results of both modalities were somehow degraded, when the user performed simultaneously the A-G commands, due to increased cognitive load. 


\begin{table}[!tb!]
\centering
\resizebox{\columnwidth}{!}{
\footnotesize{
\begin{tabular}{l|c|c|c|c|c|c||c|c|c|c|}
\cline{2-11}
& \multicolumn{6}{c||}{\textbf{System Performance \%}}& \multicolumn{4}{c|}{\textbf{User Performance \%}}\\ \cline{2-11} 
& \multicolumn{3}{c|}{\textbf{MCRR \%}} & \multicolumn{3}{c||}{\textbf{Accuracy \%}} & \multicolumn{2}{c|}{\textbf{Speech}} & \multicolumn{2}{c|}{\textbf{Gestures}} \\ \cline{2-11} 
& \textbf{L}& \textbf{B}& \textbf{Av.}& \textbf{L}& \textbf{B}& \textbf{Av.}
& \textbf{L}& \textbf{B}& \textbf{L}& \textbf{B}        \\ \hline
\multicolumn{1}{|r|}{\textbf{FSL}}       & 80& 87&83.5& 86& 73&79.5 & 98& 99& 81& 78 \\ \hline
\multicolumn{1}{|l|}{\textbf{Bethanien}} & 85& 74& 79.5&67& 77&72 & 91& 90& 84& 71 \\ \hline
\end{tabular}}}
\vspace{-0.3cm}
\caption{Average Audio-Gestural Command Recognition Results; system performance (\%) and user performance (\%) averaged across 25 and 29 users at FSL and Bethanien Hospitals, respectively. (L stands for legs, B for back and Av. for average.}
\label{fig:mcrr}
\vspace{-0.5cm}
\end{table}


\begin{figure}[!tb!]
\centering
\includegraphics[trim=0 0cm 0 0cm, clip, width=\linewidth]{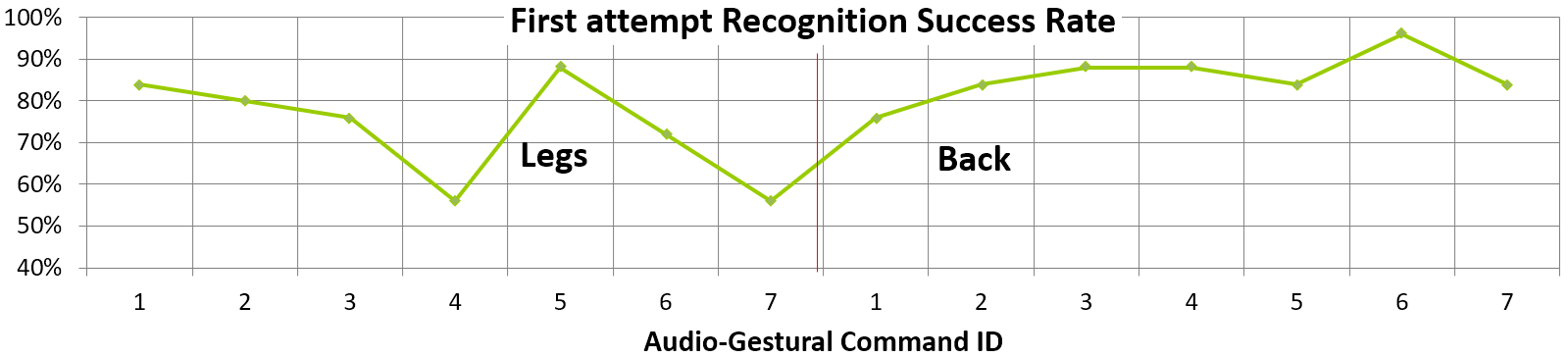}
\includegraphics[trim=0 0cm 0 0cm, clip, width=\linewidth]{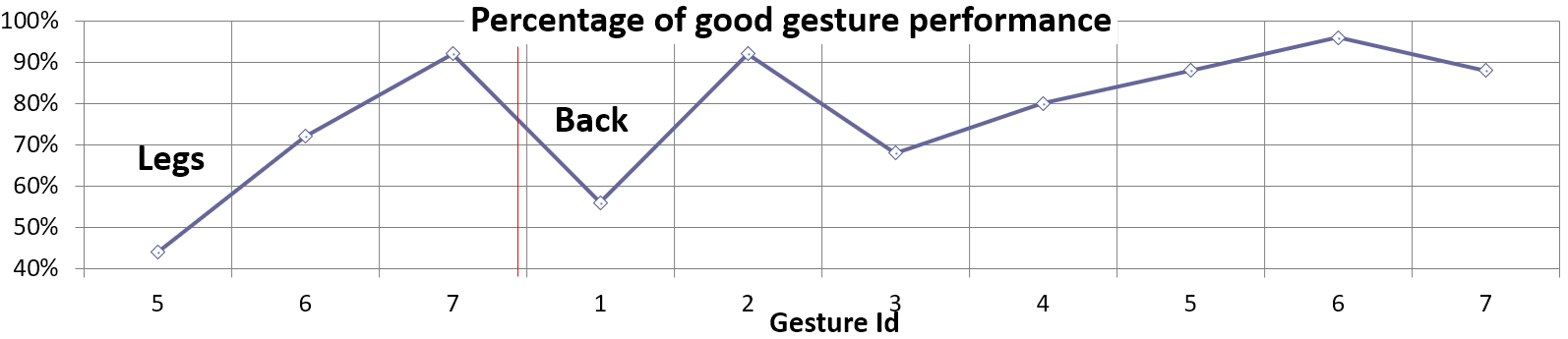}
\vspace{-0.6cm}
\caption{ Recognition statistics per command (FSL data). The vertical red lines distinguish the command sequences (see the corresponding IDs in Table~\ref{tab:IDs} for the tasks washing the legs and washing the back.}
\label{fig:recsuccrate}
\vspace{-0.65cm}
\end{figure}

Figure~\ref{fig:recsuccrate} shows indicative curves on how the users performed on their first attempt each gesture command after the training. We note that initially (gesture ID 5) the users either were not familiar with this type of communication or their concentration level was low since they were performing only spoken commands up to that point. There was however a tendency of increased learning rate, meaning that during the experiments the users got more familiar with the multimodal commands and executed them more accurately, indicating the intuitiveness of this HRI modality. Especially for commands such as ``Halt'' which was repeated several times (id 4,6,7) during the washing sequence the command performance of the user reached levels higher than 90\%. This observation is highly important, since we can conclude that simple combinations of spoken and gestural commands are both memorable and suitable for elderly user's communication with an assistive robotic system. 

\vspace{-0.1cm}
\section{Conclusions}
\label{conc}
\vspace{-0.3cm}
In this work, we presented a multimodal interface for an assistive bathing robot and a real-life use case, providing a rich set of tools and data. Such resources can be employed to develop natural interfaces for multimodal interaction between humans and robotic agents. Our intention is to further investigate how the communication between end-users and the robot will be as intuitive and effortless as possible using co-speech gesturing, which is the most natural way for human-human communication, while also enhancing the recognition, in cases of speech dysfluencies or kinetic problems. The presented online results are considered really promising given the difficulties of the task. By sharing such resources, we aim to build a public crowdsourced library that shall open new perspectives in smart assistive HRI. Regarding future work, we plan to involve the incorporation of deep learning into our online audio-gestural recognition system, 
expecting this way increased accuracy and robustness, as well as improved system response time. Finally, we intent to develop more effective audio and gesture recognition modules that will be able to make more refined decisions and process sub-sequences of gestures. 

\vspace{-0.4cm}
\section*{\footnotesize Acknowledgments}
\vspace{-0.35cm}
\footnotesize{
The authors would like to thank N. Kardaris (NTUA, Greece) for the online A-G recognition system, R. Annicchiarico and I. Griffini (FSL, Italy) and K. Hauer and C. Werner (Bethanien, Germany) for their contribution and insightful comments during the online validation of the I-Support system.}


%

\vfill\pagebreak

\label{sec:refs}

\bibliographystyle{IEEEbib}
\bibliography{acm_LateBreakingNews,action_apr2015,new}


\end{document}